\documentstyle[multicol,aps,epsf]{revtex}
\begin{document}
\title{Instability and Fluctuations of Flux Lines with
Point Impurities \\
in a Parallel Current}
\author{Mohammad Kohandel}
\address{Institute for Advanced Studies in Basic Sciences,
Zanjan 45195-159, Iran \\
Institute for Studies in Theoretical Physics and Mathematics,
Tehran 19395-1795, Iran}
\author{Mehran Kardar}
\address{Department of Physics,
Massachusetts Institute of Technology, Cambridge, Massachusetts, 02139 }
\date{\today}
\maketitle
\begin{abstract}
A parallel current can destabilize a single flux line (FL), or an array of FLs. 
We consider the effects of pinning by point impurities on this instability. 
The presence of impurities destroys the long-range order of a flux lattice, 
leading to the so called Bragg glass (BrG) phase.
We first show that the long-range topological order of the 
BrG is also destroyed by a parallel current. 
Nonetheless, some degree of short-range order should remain , 
whose destruction by  thermal and impurity fluctuations, 
as well as the current, is studied here.
To this end, we employ a cage model for a single FL in the presence o
f impurities and current, and study it analytically (by replica variational 
methods), and numerically (using a transfer matrix technique). 
The results are in good agreement, and in conjunction with a 
Lindemann criterion, provide the boundary in the 
magnetic field--temperature plane for destruction of short-range order. 
In all cases, we find that the addition of impurities or current (singly or
in combination) leads to further increase in equilibrium FL fluctuations.
Thus pinning to point impurities does not stabilize FLs in a parallel current $j_z$, 
although the onset of this instability is much delayed due to large
potential barriers that diverge as $j_z^{-\mu}$.

\vspace{1cm}
\end{abstract}
\begin{multicols}{2}

\section{Introduction and Summary}

There is currently much effort devoted  to understanding the nature of
vortex phases in high temperature superconductors \cite{Blatter}.
The flux lines (FLs) in a clean material prefer to sit on the sites of a
triangular lattice forming an Abrikosov solid for fields $H>H_{c_1}$.
However, this mean field phase diagram is modified by the combined
effects of thermal fluctuations and various types of disorder, such
as oxygen vacancies, columnar defects, or twin boundaries.
Understanding the precise nature of these new phases is very
important for the determination of the transport properties of these
materials. In this paper, we focus on the combined effects of 
point-like disorder, such as oxygen vacancies, and a parallel current. 
(There are many studies that consider other
types of disorder \cite{Nelson,Balent}).

It is known that pinning to even a small amount of random impurities
is sufficient to destroy the long-range translational order of
the Abrikosov lattice\cite{Larkin}. If  randomness is strong, the underlying
order is destroyed at every scale, but for weak enough randomness, 
some order is expected to survive at short distances.
Indeed, it has been suggested that there may exist a stable, 
dislocation-free Bragg Glass (BrG) phase at low magnetic fields
and temperatures in the presence of weak point impurities \cite{GL,EN}.
It is important, both for technical applications of high $T_c$
materials and from a theoretical point of view, to understand 
the stability of this new phase with respect to an external current.
For example, one may wish to control the melting transition
of the BrG phase by some external parameter such as current.
The case of a current perpendicular to the vortex lines was
studied in Ref.~\cite{GL2}, where it was shown that the driven
lattice may still maintain a topologically ordered moving BrG phase.
Here, we study the case in which  current is applied parallel to the FLs.
The instability of the proposed equations for the BrG suggests
that the long-range topological order of this phase is most likely
destroyed by a parallel current.
Using a Lindemann criterion and a cage model, we then discuss the 
destruction of any remnant short range order.
This is in the same spirit as a recent calculation\cite{MKMK},
in which we used a Langevin equation to examine
the dynamics and fluctuations of a FL in the presence of
an {\em alternating} parallel current.
The basic idea is that even in the absence of true long-range order
due to current or impurities,
there remains a great deal of translational order in the vortex system
at short scales. 
A ``phase diagram" is then constructed to indicate the apparent limits of
this ``ordered" region.

The response of FLs to a parallel current $j_{z}$ was studied 
extensively even prior to the high-$T_c$ era \cite{Brandt1}.
While, a straight FL is stable in the presence of $j_{z}$,
any fluctuation of the line causes helical instabilities
at long wavelengths \cite{Brandt1}.
A lattice of FLs is similarly unstable in the
presence of an infinitesimal
$j_{z}$, if it penetrates the whole sample. As shown by Brandt
\cite{Brandt2}, for a finite penetration depth of current,
a finite $j_{z} \sim {\varepsilon}^{{1}/{4}}$, is
needed to cause instability, where $\varepsilon$ is related to the shear
modules $c_{66}$. Brandt has suggested \cite{Brandt1}  that FLs can
also be stabilized for infinitesimal
$j_{z}$ (which decays exponentially into the sample)
due to pinning by disorder.
However, this study predates the
introduction of the BrG phase, which is more ordered than a flux liquid.
We may thus ask if the BrG phase is stable with respect to $j_{z}$?
More generally, what is the behavior of the flux line in response to 
the current $j_{z}$ when point impurities are present?
Since, the melting of the BrG phase is thought to be first order,
the possibility of diverging
longitudinal and perpendicular penetration depths
of the current at the transition point is excluded.
Nonetheless, we shall assume a current $j_{z}$ which
uniformly penetrates the whole sample, and calculate the
mean-squared fluctuations of the FL 
based on this assumption \cite{FotN,MKMK}.
This quantity is then used to examine the extent 
of short-range order, and to construct an apparent 
``phase diagram" in the magnetic field--temperature ($B-T$) plane.

Two models used for the study of FLs with disorder are introduced
in Sec.~\ref{models}. 
The first model describes the elastic behavior of  the vortex lattice
in the presence of impurities, and is appropriate for the study of 
its long-wavelength fluctuations.  
The true long-range order of the lattice is destroyed by such fluctuations,
and possibly replaced by the weaker topological order of the proposed
Bragg glass\cite{GL}.
The second model considers a single FL confined by a harmonic cage
potential that represents interactions with the other FLs.
This model provides an approximate measure of the extent of short-range
fluctuations, and in conjunction with a Lindemann criterion can be used
to construct a phase diagram\cite{EN}. 
The addition of current to these models is straightforward.

In Sec.~\ref{replicas} replica variational methods are used to estimate
FL fluctuations in the presence of impurities and current.
The BrG phase is described by a variational ansatz that breaks full 
replica symmetry. 
We propose a natural extension of this solution in the presence of current.
This solution, however, has unbounded fluctuations and suffers from
the same long-wavelength instabilities as the uniform lattice.
We interpret this as a signal of the destruction of BrG order by the current.
(Naturally, we cannot rule out other potential variational solutions
with other forms of remnant long-range order.)
However, it is quite likely that some degree of short-range translational
order survives for weak currents and impurities.
Using the cage model, we then estimate the fluctuations of a single FL.
To this end, we generalize the replica variational ansatz proposed
by Goldschmidt\cite{Gold} to include a current. 
The mean-squared fluctuations of the FL are found to
grow with increasing current and/or impurities.
Consequently, in the absence of the cage potential,
a single FL is unstable in the presence of an infinitesimal current, 
and is not stabilized by the impurities alone.

We test the predictions of the replica calculations (for the single FL in a cage)
using a transfer matrix method which is introduced in Sec.~\ref{tm}.
For a given realization of randomness, this method provides 
numerically {\em exact} results for the extent of fluctuations of a
discretized FL in a cage potential.
While it is difficult to provide an exact correspondence with the corresponding
continuum limit, the results are qualitatively in agreement with
those of the replica variational ansatz.
In particular, we again find that both impurities and currents enhance
transverse fluctuations, and that the  FL is unstable in the absence of 
the cage potential.

The conclusion that  pinned FLs are unstable, even to infinitesimal
current $j_z$, may appear as somewhat counterintuitive:
The force due to the current on a portion of the FL is then much less
than the attractive potential from impurities.
The resolution lies in the size of the energy barriers that separate 
metastable pinned configurations.
In Sec.~\ref{onset} we argue that the size of these barriers diverges
as $j_z^{-\mu}$, where the exponent $\mu$ is related to 
collective pinning effects.
At finite temperatures, an exponentially long time is required for
activated hopping over these barriers.
Thus the onset of the instability can be much delayed due to pinning 
by impurities.

\section{Model Hamiltonians}\label{models}

\subsection{Flux lattice with impurities}
To discuss the large-scale deformations of the Bragg glass,
we employ the elastic description introduced  
by Giamarchi and Le Doussal \cite{GL}.
If the displacements of the FLs with respect to their equilibrium positions,
$\vec R_i$, are denoted by the two-component 
displacement field $\vec u(\vec R_i,z)$, then the energy cost of deforming the 
lattice, in the continuum limit, is given by
\begin{eqnarray}\label{HamL}
{\cal H}_{L}&=&\frac{1}{2}\sum_{\alpha\beta}\int \frac{d^3q}{(2\pi)^3}u_\alpha(q)
\Phi_{\alpha\beta}(q)u_\beta(-q)   \\
&+&\int d^3xV(x)\rho(x),  \nonumber
\end{eqnarray}
where $\alpha,\beta=1,2$ label the coordinates.
The first term in the above equation denotes the elastic energy,
and the $q$--integration is over the Brillouin zone. 
The symmetries of the lattice constrain the form of the elastic kernel
$\Phi_{\alpha\beta}$, for example resulting in two elastic moduli as
$q\to0$ in an isotropic system.
For the sake of brevity, following Ref.~\cite{GL}, we shall assume
the simplified form of $\Phi_{\alpha\beta}(q)=cq^2\delta_{\alpha\beta}$.
The second term in Eqn.~(\ref{HamL}) is related to the potential 
produced by the point impurities.
In the limit where many weak impurities act collectively on a FL, the
resulting disorder is modeled by a Gaussian random potential with 
correlations $\overline{V(x)V(x')}=\Delta(|\vec r-\vec r'|)\delta(z-z')$. Here,
$\vec r$ and $z$ denote coordinates perpendicular and parallel
to the FL respectively, while
$\Delta(r)$ is a short range function of the scale of the superconducting
coherence length. The density of the vortices at a given point $x$
is also given by
\begin{equation}\label{rhox}
\rho(x)=\sum_i \delta\left(\vec r-\vec R_i-\vec u(\vec R_i,z)\right).
\end{equation}

The average over the quenched disorder can be performed using
the replica method.
After averaging over disorder, the $n$-th moment of the partition 
function at a temperature $T$
is the same as that of the replicated Hamiltonian \cite{GL},
\begin{eqnarray}\label{Heff1}
{\cal H}_{eff}&=&\frac{c}{2}\int d^3x\left(\partial_\alpha u^a_\beta\right)^2-
\int d^3x\sum_{a,b}\Bigg[\frac{\Delta_0}{2T}
\partial_\alpha u^a_\alpha\partial_\beta u^b_\beta  \nonumber  \\ 
&+&\sum_{\vec K\neq 0}\frac{\Delta_K}{2T}\cos\left\{\vec K\cdot
\left[\vec u^a(x)-\vec u^b(x)\right] \right\}\Bigg],
\end{eqnarray}
where $a,b=1,\cdots,n$ refer to the replica indices.
Due to the lattice periodicity, averaging over disorder introduces
both harmonic ($\Delta_0$) and anharmonic ($\Delta_K$) interactions
between the replicated fields, where $\vec K$ is 
any reciprocal lattice vector, and $\Delta_K$ is the Fourier
component of $\Delta(r)$.

Generalizing the above Hamiltonians to include a current $j_z$,
parallel to the magnetic field, is straightforward, and achieved
by adding the term $\phi_0j_z\vec u\times \partial_z\vec u/2$.
In general the current $j_z$ may be inhomogeneous, decaying
from the surface into the bulk. 
To further simplify the calculations, we shall assume that the
current uniformly penetrates the sample\cite{FotN}.
This assumption enables us to write the replicated Hamiltonian as
\begin{eqnarray}\label{Heff2}
{\cal H}_{eff}&=&\frac{1}{2}\int \frac{d^3q}{(2\pi)^3}\sum_a
G_{0,\alpha\beta}^{-1}(q)u^a_\alpha(q) u^a_\beta(-q) \nonumber \\   
&-&\int d^3x\sum_{a,b}\Bigg[\frac{\Delta_0}{2T}
\partial_\alpha u^a_\alpha\partial_\beta u^b_\beta  \\  \nonumber
&+&\sum_{K\neq 0}\frac{\Delta_K}{2T}
\cos\left\{\vec K.\left[\vec u^a(x)-\vec u^b(x)\right]\right\}\Bigg],
\end{eqnarray}
where $G_{0,\alpha\beta}^{-1}(q)=cq^2\delta_{\alpha\beta}+
i\phi_0 j_z q_z\epsilon_{\alpha\beta}$.
Properties of this Hamiltonian will be discussed in Sec.\ref{replicaA}.

\subsection{The Cage Model for a single FL}

To examine fluctuations of a representative FL at short scales, 
we shall use  the {\it cage model}.
The collective interactions of the other FLs is then approximated by a 
simple harmonic potential. 
The applicability of the cage model to the BrG
is discussed in Refs.~\cite{EN,MKMK}.
We again consider the limit where many weak impurities
act collectively on the FL, so that the disorder can be modeled by a
Gaussian random potential. 
Thus, the Hamiltonian for a single FL, whose position is denoted
by the two-component vector $\vec r\,(z)$, is\cite{MKMK,Gold}
\begin{eqnarray}\label{Ham}
{\cal H}=\int dz \Bigg[\frac{\varepsilon_1}{2} \left(\frac{\partial \vec r}
{\partial z}\right)^2+\frac{\mu}{2}\,\vec r\,^2+V\left(z,\vec r\,\right)+ 
\nonumber \\
\frac{\phi_0 j_z}{2}\vec r\,\times\partial_z\vec r\Bigg].
\end{eqnarray}
Here, $\varepsilon_1=\varepsilon_0/\gamma^2$ is the line tension
of the FL, $\gamma^2=m_z/m_{\perp} \gg 1$ is the mass anisotropy,
$\varepsilon_0=(\phi_0/{4 \pi \lambda})^2$, and $\phi_0=hc/2e$.
The spacing of the FLs is indicated by $a$, while $\lambda$
is the penetration depth \cite{Blatter}. The spring constant $\mu$,
is approximated in the dense limit of $B>H_{c_1}$, by $\mu \approx
{\varepsilon_0}/{a^2}$. The last term of the Hamiltonian
is generated by the Lorentz force due to the parallel current.
The random potential is
assumed to have the following correlations\cite{EN,Gold},
\begin{equation}\label{randomp}
\left\langle V(z,\vec r)V(z^\prime,\vec r\,^\prime)\right\rangle=
\tilde \Delta\epsilon_0^2 \xi^3 \delta_{\xi}^{(2)}(\vec r-
\vec r\,^\prime)\delta(z-z^\prime),
\end{equation}
where
\begin{equation}\label{delta}
\delta_{\xi}^{(2)}(\vec r-\vec r\,^\prime)\approx 1/(2\pi\xi^2)
\exp\left[-(\vec r-\vec r\,^\prime)^2/2\xi^2\right].
\end{equation}
The dimensionless parameter $\tilde \Delta$ is
a measure of the strength of the disorder, and $\xi$
is the vortex core diameter (coherence length).

Goldschmidt \cite{Gold} has used replica and variational 
methods to compare and contrast the effects of point and
columnar disorder in the above Hamiltonian.
In Sec.\ref{replicaB} we shall generalize his results to the case
of a non-zero current.

\section{The Replica variational method}\label{replicas}

\subsection{Instability of the Bragg Glass}\label{replicaA}

In the absence of disorder, the elastic Hamiltonian of Eqn.~(\ref{Heff2})
has normal mode energies of $c q^2\pm \phi_0j_zq_z$. 
Clearly, there are negative energies at small $q_z$ for any finite
$j_z$, indicating the instability of the lattice to long wavelength deformations.
We ask if a similar instability is present in the presence of disorder,
when the vortex lattice is replaced by a Bragg glass.
To this end, we shall first briefly review the approach of
Giamarchi and Le Doussal \cite{GL}, and then generalize it to a finite $j_z$. 

It is convenient to first decompose the 
propagator $G^{-1}_{0,\alpha\beta}$ as,
\begin{eqnarray}\label{DecomG}
G^{-1}_{0,\alpha\beta}&=&(cq^2+\phi_0j_zq_z) 
[(\delta_{\alpha\beta}+i\varepsilon_{\alpha\beta})/2]    \\  \nonumber
&+&(cq^2-\phi_0j_zq_z)
[(\delta_{\alpha\beta}-i\varepsilon_{\alpha\beta})/2] \\  \nonumber 
&=&G_0^{-1,L}P^L_{\alpha\beta}+G_0^{-1,R}P^R_{\alpha\beta},
\end{eqnarray}
where $P^{L,R}_{\alpha\beta}$ are  projection operators 
(to left and right circular polarizations), which satisfy $P^i_{\alpha\beta}P^j_{\beta\gamma}=\delta_{ij}
\delta_{\alpha\gamma}$.

The essence of the variational approach is to find an optimal
Gaussian trial Hamiltonian of the form
\begin{equation}\label{H02}
{\cal H}_0=\int \frac{d^3q}{2(2\pi)^3} G^{-1}_{\alpha\beta,ab}(q) 
u^a_\alpha(q)u^b_\beta(-q),
\end{equation}
where $G^{-1}_{\alpha\beta,ab}=
G^{-1,L}_{ab}(q)P^L_{\alpha\beta}+G^{-1,R}_{ab}(q)P^R_{\alpha\beta}.$
The generic form chosen for the trial propagators is 
$G^{-1,i}_{ab}=\delta_{ab}G^{-1,i}_0-\sigma_{ab}^{i}$.
The matrices $\sigma^{L/R}_{ab}$ act as effective caging potentials 
due to disorder, and may in principle stabilize the system.
The best variational approximant to the Hamiltonian of Eqn.~(\ref{Heff2})
is obtained by minimizing the variational free energy 
$F=F_0+\left\langle {\cal H}_{eff}-{\cal H}_0\right\rangle$,
and results in 
\begin{equation}\label{sigmaL}
\sigma^{L/R}_{a\neq b}=\sum_K \frac{\Delta_K}{2T}K^2
\exp\left[ -\frac{K^2}{4}\left(B_{ab}^L+B_{ab}^R\right)\right],
\end{equation}
where 
\begin{eqnarray}\label{BabL}
B_{ab}^{L/R}=T\int \frac{d^3q}{(2\pi)^3}\Bigg[ G_{aa}^{L/R}(q)+G_{bb}^{L/R}(q)
\nonumber  \\
-2G_{ab}^{L/R}(q)\Bigg].
\end{eqnarray}

These equations have been discussed extensively in Ref.~\cite{GL}.
Remarkably, they have solutions which break the replica symmetry.
The full breaking of replica symmetry is best represented by replacing
the symbols ${ab}$ with a continuous variable $v\in (0,1)$.
The previous equations for matrix elements can now be written
as coupled integral equations
\begin{eqnarray}\label{sigmaBabL}
\sigma^{L/R}(v)&=&\sum_K\frac{\Delta_K}{2T} K^2\exp\left[
-\frac{K^2}{4}\left(B^L(v)+B^R(v)\right)\right], \nonumber  \\
B^{L/R}(v)&=&2T\int \frac{d^3q}{(2\pi)^3}\left[\tilde G^{L/R}(q)-
G^{L/R}(q,v)\right].
\end{eqnarray}
(The propagators $G_{ab}(q)$ and 
$G_{aa}(q)$ are replaced by $G(q,v)$ and $\tilde G(q)$ respectively.) 
It is apparent from Eqs.~(\ref{BabL}) and (\ref{sigmaBabL}) that
$\sigma^L(v)=\sigma^R(v)$, and we shall henceforth drop the indices $L/R$.
A typical solution for $\sigma(v)$ is a constant for $v>v_c$, 
and an arbitrary function for $v<v_c$, where $v_c$ is itself 
a variational parameter.
The algebraic rules for inversion of hierarchical matrices give\cite{GL,MezPar},
\begin{eqnarray}\label{BLT2}
B^{L/R}(v)&=&B^{L/R}(v_c)  \\  
&+&\int_v^{v_c}d\omega\int\frac{d^3q}{(2\pi)^3}
\frac{2T\sigma^\prime(\omega)}
{\left(G_0^{-1,L/R}(q)+[\sigma](\omega)\right)^2}, \nonumber
\end{eqnarray}
where $\sigma^\prime(v)=d\sigma/dv$, 
$[\sigma](v)=v\sigma(v)-\int_0^vd\omega\sigma(\omega)$, and
\begin{equation}\label{BLTvc2}
B^{L/R}(v_c)=\int\frac{d^3q}{(2\pi)^3}
\frac{2T}{G_0^{-1,L/R}(q)+[\sigma](v_c)}.
\end{equation}

To simplify calculations, we keep only the first term, with $K=K_0$,
in the sum in Eqn.~(\ref{BabL}). 
(This is the single cosine model of the Ref.\cite{GL}).)
Eqs.~(\ref{BabL})-(\ref{BLTvc2}) are now solved by first differentiating
the expression for $\sigma(v)$ with respect to $v$.
Eliminating $\sigma^\prime(v)$ from both sides of the equation,
and performing the integrals over $q$, yields
\begin{equation}\label{Rsigmav1}
{\sigma(v)}=\frac{8\pi c^{3/2}}{K_0^2T}{\sqrt{[\sigma](v)-(j_z/j_0)^2}} , 
\end{equation}
for $[\sigma]>(j_z/j_0)^2$, and $\sigma(v)=0$ for $[\sigma]<(j_z/j_0)^2$, 
where $j_0=\sqrt{4c/\phi_0^2}$. 
Taking one more derivative, and using $[\sigma]'(v)=v\sigma'(v)$,
gives
\begin{equation}\label{Rsigmav2}
[\sigma](v)=\left(\frac{v}{v_0}\right)^2+\left(\frac{j_z}{j_0}\right)^2,
\end{equation}
where $v_0=K_0^2 T/(4\pi c^{3/2})$.

Note that the above solution for $[\sigma](v)$ is always positive,
and hence incompatible with the initial condition of $[\sigma](v=0)=0$.
A potential resolution\footnote{Suggested to us 
by M. Mezard in private communications.}
is to assume that it goes to zero  {\em discontinuously} 
for $v$ less than some $v_c'$, i.e. 
\begin{equation}\label{sigmavc}
[\sigma](v)=\left\{
\begin{array}{ll}
0 &{\rm,~for~}  v<v_c', \\
\left(\frac{v}{v_0}\right)^2+\left(\frac{j_z}{j_0}\right)^2
&{\rm,~for~} v_c'<v<v_c, \\   
\Sigma &{\rm,~for~} v>v_c.
\end{array}
\right.
\end{equation}
Self-consistency of this solution with the condition 
$[\sigma](v)=v\sigma(v)-\int_0^vd\omega\sigma(\omega)$
requires $v_c'=v_0(j_z/j_0)$, and thus gives
$[\sigma](v=v_c'^+)=2(j_z/j_0)^2$.
We can assume that $[\sigma]$ is continuous at $v_c$, 
leading to the condition $v_c=v_0\sqrt{\Sigma-(j_z/j_0)^2}$.
Another relation between these two unknowns is obtained
by substituting $v=v_c$ in Eqn.~(\ref{sigmaBabL}), leading to
\begin{eqnarray}\label{sigvc}
\sqrt{\Sigma-{j_z^2\over j_0^2}}=\frac{\Delta_0 K_0^4}{16\pi c^{3/2}}
\exp{\left[-\frac{K_0^2}{4}\left(B^L(v_c)+B^R(v_c)\right)\right]}.
\end{eqnarray}
where $B^{L/R}(v_c)$ is defined in Eqn.~(\ref{BLTvc2}), with 
$[\sigma](v_c)=\Sigma$. 
Solutions to these equations (for $j_z=0$) are extensively
discussed in the Ref.~\cite{GL}. 

We shall now inquire if these solutions to the variational equations
are physically sensible.
In particular, the variance of fluctuations at a wavenumber $q$ is
controlled by the correlation functions,
\begin{equation}\label{FinG1}
\tilde G^{L/R}(q)=G_0^{L/R}(q)\left[1+\int_0^1\frac{dv}{v^2}\frac{[\sigma](v)}
{\left(G_0^{-1,L/R}(q)+[\sigma](v)\right)}\right].
\end{equation}
Using the expression for $[\sigma](v)$, and performing the integrals
over $v$, leads to 
\begin{eqnarray}\label{FinG2}
\tilde G^{L/R}(q)&=&\frac{1}{cq^2\pm \phi_0j_z q_z}\Bigg[1+\frac{c_1}
{cq^2\pm \phi_0j_z q_z+(j_z/j_0)^2}  
\nonumber  \\
&+&\frac{c_2(cq^2\pm \phi_0 j_z q_z)}{\left(cq^2\pm \phi_0j_z q_z
+(j_z/j_0)^2\right)^{3/2}}
\nonumber  \\
&+&\frac{1-v_c}{v_c}\frac{\Sigma}{cq^2\pm \phi_0j_zq_z+\Sigma}\Bigg],
\end{eqnarray}
where $c_1=(1/v_c'-1/v_c)(j_z/j_0)^2$, and
$c_2=[\tan^{-1}(v_c/v_c')-\pi/4]/v_0$.
For $j_z=0$, the leading behavior as $\vec q\to 0$ comes from the third
term in Eqn.~(\ref{FinG2}), and results in $\tilde G^{L/R}\sim 1/|q|^3$.
This is the characteristic signature of the BrG, quite distinct from the
$1/q^2$ scaling of an elastic material in the absence pinning, 
and leads to logarithmic growth of fluctuations in real space\cite{GL}.
The modified singularity can be traced back to the $v\to0$ limit of 
the integral in Eqn.~(\ref{FinG1}).
A finite current removes this singularity as indicated by Eqn.~(\ref{sigmavc}),
and in the limit of $\vec q\to 0$ the integrals evaluated in the square
brackets of Eqn.~(\ref{FinG2}) go to a constant.
Thus, apart from a multiplicative constant, the form of the correlator is
preserved, and the solution suffers from the same instabilities as
in the absence of pinning.
(In other words, our Gaussian variational solution has some eigenmodes
with negative eigenvalues.)
We therefore conclude that this natural extension of the BrG in the presence
of current is not stable.
While it is  possible that there are other variational solutions that do not
suffer such instability, the most likely implication is that the (quasi-)long-range
order of the BrG is destroyed by the parallel current.
 
\subsection{A single FL}\label{replicaB}

The mean-square  fluctuations of each FL (due to short-wavelength
fluctuations) are better estimated using the cage Hamiltonian of Eqn.~(\ref{Ham}).
The average over the quenched random potential is again performed
by introducing $n$ copies of the FL. After averaging over disorder,
the replicated displacements (in Fourier space)
of the FLs are coupled through the Hamiltonian
\begin{eqnarray}\label{HamN}
{\cal H}_n&=&\int \frac{dq_z}{2(2\pi)}\,\sum_{a=1}^n
G^{-1}_{0,\alpha\beta}(q_z) r^a_\alpha(q_z)r^a_\beta(-q_z) \\ \nonumber
&-&\frac{\beta\tilde\Delta\xi\epsilon_0^2}{4\pi}\,\int dz
\sum_{a,b} \exp\left[-\frac{(\vec r\,_a-\vec r\,_b)^2}{2\xi^2}\right],    
\end{eqnarray}
where
\begin{equation}\label{G0-1}
G^{-1}_{0,\alpha\beta}(q_z)=\left(\varepsilon_1 q_z^2+\mu\right)
\delta_{\alpha\beta}+iq_z\phi_0 j_z \varepsilon_{\alpha\beta}.
\end{equation}
We again search for the best variational Gaussian Hamiltonian $h_n$,
parameterized as
\begin{equation}\label{hn}
h_n=\int \frac{dq_z}{2(2\pi)}\,G^{-1}_{ab,\alpha\beta}(q_z)
r^a_\alpha(q_z) r^b_\beta(-q_z).
\end{equation}
Once more, it is convenient to decompose the propagator in Eqn.~(\ref{G0-1}) 
into a sum of left and right circular polarizations as in Eqn.~(\ref{DecomG}), where  
$G_0^{-1,L/R}=(\varepsilon_1 q_z^2\pm \phi_0 j_z q_z+\mu)$.
In terms of the projection operators, the inverse of the propagator 
for the trial Hamiltonian is simply
\begin{equation}\label{Gabt}
G_{ab,\alpha\beta}=G_{ab}^L(q_z)P^L_{\alpha\beta}+G_{ab}^R(q_z)
P^R_{\alpha\beta}.
\end{equation}

The trial propagator  is written as 
$G^{-1}_{ab,\alpha\beta}(q_z)=\delta_{ab} G^{-1}_{0,\alpha\beta}(q_z)-
\sigma_{ab,\alpha\beta}$, with $\sigma_{ab,\alpha\beta}$ determined by 
minimizing the variational free energy
\begin{equation}\label{FE1}
F=\left\langle {\cal H}_n-h_n\right\rangle_{h_n}-T
\log\left(\int d[\vec r\,_a]e^{- h_n/T}\right).
\end{equation}
Using Eqs. (\ref{HamN}), (\ref{hn}) and (\ref{FE1}), and after
some algebraic calculations one finds the free energy 
\begin{eqnarray}\label{FE2}
&&\frac{F}{2L}=
{\rm constant}+\frac{T}{2}\int \frac{dq_z}{2\pi}\sum_{\alpha,\beta,a}
G^{-1}_{0,\alpha\beta}(q_z) G_{aa,\alpha\beta}(q_z)  \nonumber \\ 
&&-\frac{T}{2}\int \frac{dq_z}{2\pi} {\rm tr}\ln G(q_z)+
\frac{1}{2T}\sum_{a,b} \tilde f_p\left(\sum_\alpha
B_{ab,\alpha\alpha}\right), 
\end{eqnarray}
where the function $\tilde f_p$ is given by
\begin{equation}\label{func}
\tilde f_p(y)=-\frac{\tilde \Delta \xi^3\varepsilon_0^2}{4\pi}
\frac{1}{\xi^2+y},
\end{equation}
and
\begin{equation}\label{Babalbe}
B_{ab,\alpha\beta}=T \int \frac{dq_z}{2\pi}
\left[G_{aa,\alpha\beta}+G_{bb,\alpha\beta}-2 G_{ab,\alpha\beta}\right].
\end{equation}
The stationary solution for the free energy is obtained for 
$\sigma_{ab,\alpha\beta}=\delta_{\alpha\beta}\sigma_{ab}$, with
\begin{eqnarray}\label{sigma}
&\sigma&_{ab}= \frac{2}{T}\tilde f_p^\prime\left[\frac{1}{2}\left(B_{ab}^L
+B_{ab}^R\right)\right], \qquad {\rm for} \qquad  a\neq b
\nonumber \\ 
&\sigma&_{aa}=\sum_{b(\neq a)}\sigma_{ab}.  
\end{eqnarray}
Here, $\tilde f_p^\prime$ is the derivative of $\tilde f_p$ with
respect to its argument, and $B^{L/R}_{ab}$ are given by
\begin{equation}\label{BabLT}
B^{L/R}_{ab}=T\int \frac{dq_z}{2\pi}\left[G^{L/R}_{aa}+G^{L/R}_{bb}
-2G^{L/R}_{ab}\right].
\end{equation}

We now proceed to discuss the solution of the stationarity
equations with the assumption of replica symmetry (RS), i.e. with
$\sigma_{ab}=\sigma$ for all off-diagonal elements. In the
limit of $n\rightarrow 0$ one finds,
\begin{equation}\label{RSS}
G^{L/R}_{ab}=G^{L/R}_0\delta_{ab} +\sigma \left(G^{L/R}_0\right)^2.
\end{equation}
Using Eqs. (\ref{BabLT}) and (\ref{RSS}) and integrating over $q_z$
we obtain
\begin{equation}\label{BabRS}
B^{L/R}_{a\neq b}=\frac{2T}{\sqrt{4\varepsilon_1\mu-\phi_0^2 j_z^2}},
\qquad {\rm for} \qquad j_z<j_c
\end{equation}
and no solution for $j_z>j_c$, where
$j_c=\sqrt{4\varepsilon_1\mu /\phi_0^2}$. 
(Note that in performing the integrals over $q_z$ 
from $-\infty$ to $+\infty$, we implicitly assume a lattice spacing
smaller than the length $\sqrt{\mu/\varepsilon_1}$.)
From Eqn.~(\ref{sigma}), the value of $\sigma$ is obtained as
\begin{equation}\label{sigmaRS}
\sigma=\frac{2}{T}\tilde f_p^\prime\left(\frac{2T}
{\sqrt{4\varepsilon_1\mu-\phi_0^2 j_z^2}}\right) .
\end{equation}
It is interesting to note that in the limit of $\mu \rightarrow 0$,
$\sigma$ is equal to zero which means that there is no influence
of disorder in the RS solutions in the absence of the cage potential.

The transverse excursions of a long FL are given by
\begin{equation}\label{CFRS}
\overline{\langle\vec r\,^2\rangle}=T\int {dq_z\over (2\pi)} 
\left[G^L_{aa}+G^R_{aa}\right].
\end{equation}
Using Eqs.~(\ref{CFRS}) and (\ref{RSS}) and integrating over $q_z$ gives
\begin{equation}\label{ICFReS1}
\overline{\langle\vec r\,^2\rangle}=\frac{(T/\sqrt{\varepsilon_1\mu})}{
\sqrt{1-(j_z/j_c)^2}}\left[
1+\frac{\sigma}{2\mu\left(1-(j_z/j_c\right)^2)}\right]  .
\end{equation}
It is convenient to introduce dimensionless parameters
\begin{equation}\label{DLP}
\tilde T=\frac{T}{\varepsilon_0\xi}, \qquad{\rm and}
\qquad \tilde B=\frac{B\xi^2}
{\phi_0},
\end{equation}
in terms of which we have $\mu=\varepsilon_0\tilde B/\xi^2$
and $a^2=\xi^2/\tilde B$. Using Eqs.~(\ref{sigmaRS})
and (\ref{DLP}), and rewriting Eqn.~(\ref{ICFReS1}) in terms of these
dimensionless parameters, we obtain
\begin{eqnarray}\label{ICFRS2}
\frac{\overline{\langle\vec r\,^2\rangle}}{a^2}
&=&\frac{\gamma \tilde T \sqrt{\tilde B}}{
\sqrt{1-(j_z/j_c)^2}} \\  \nonumber
&+&\frac{\gamma\tilde \Delta}{4\pi\sqrt{\tilde B}}
\frac{\left[1-(j_z/j_c)^2\right]^{-3/2}}{\left\{1+\gamma\tilde T\left[{\tilde B}
\left(1-(j_z/j_c)^2\right)\right]^{-1/2}\right\}^2}.
\end{eqnarray}
Since in the real system we have $\tilde B\ll(\gamma\tilde T)^2$,
Eqn.~(\ref{ICFRS2}) can be further simplified to
\begin{equation}\label{ICFRS3}
\frac{\overline{\langle\vec r\,^2\rangle}}{a^2}=
\frac{\gamma \tilde T \sqrt{\tilde B}}{
\sqrt{1-(j_z/j_c)^2}}\left[1+\frac{\gamma\tilde \Delta}{4\pi}
\frac{1}{(\gamma \tilde T)^3}\right] .
\end{equation}
As expected, this relation predicts that the mean square 
displacements of the FL grow with increasing disorder and/or current.

The mean square fluctuations from Eqn.~(\ref{ICFRS3}) are plotted
in Figs.~1(b) and 1(d), for zero and finite disorder, with the parameters 
$\gamma=125$, $\xi=30\AA$, $\varepsilon_0\xi=1905^\circ K$, 
at $T=50^\circ K$, and with $B=250G$.
(Figures 1 (a) and 1(c), at the lower temperature of $T=30^\circ K$
require the use of 1-step RSB, which will be discussed shortly.)
These are the same set of parameters used by Goldschmidt in Ref.~\cite{Gold}.
Indeed, we have essentially followed his approach and notation
throughout this section, generalizing to a finite $j_z$.
There is, however, one point of difference in calculating the extent of
transverse fluctuations of the FL:
We use Eqn.~(\ref{CFRS}) which sums over contributions of all modes.
Goldschmidt\cite{Gold} considers fluctuations of a FL segment of length 
$l^*=\sqrt{\epsilon_1/\mu}$, which is similar to removing the contributions
for $q_z<1/l^*$ from Eqn.~(\ref{CFRS}). 
Thus, for the same set of parameters, we obtain larger fluctuations
(and a lower melting temperature).

\begin{figure}
\epsfxsize=7truecm
\centerline{\epsfbox{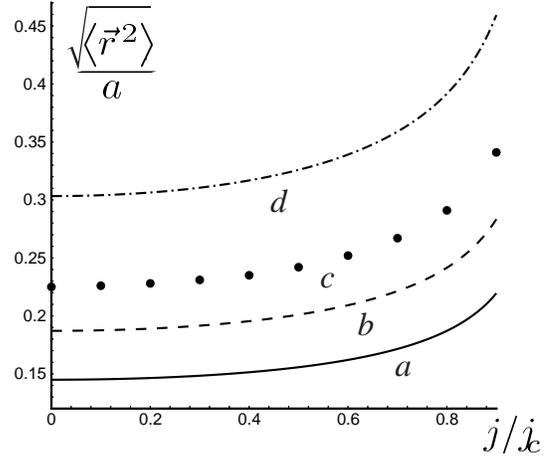}} 
\smallskip
\narrowtext{\caption{The mean-square fluctuations of a FL with zero disorder
($\tilde \Delta=0$) for (a) $T=30^\circ K$, (b) $T=50^\circ K$; 
and finite disorder ($\tilde \Delta/2\pi=0.2$) for (c) $T=30^\circ K$, 
(d) $T=50^\circ K$. \label{fig1} }}
\end{figure}

The next step is to look for variational solutions which break replica symmetry.
The technical details of this calculation are presented in the appendix.
Following Goldschmidt\cite{Gold}, we find a solution with 1-step replica 
symmetry breaking (RSB) for temperatures below a $j_z$-independent
freezing transition point of
\begin{equation}
T_{cp}\approx (\varepsilon_0\xi/\gamma)(\gamma\tilde\Delta/2\pi)^{1/3}.
\end{equation}
The mean squared fluctuations with the 1-step RSB, again in the limit of
$\tilde B\ll(\gamma\tilde T)^2$, are obtained from
\begin{eqnarray} \label{flu1srsb}
&&\frac{\overline{\langle \vec r\,^2\rangle}}{a^2}=\gamma
\tilde T\sqrt{\tilde B}\Bigg\{\frac{1}{v_c\sqrt{1-(j_z/j_c)^2}}-\left(\frac{1-v_c}{v_c}\right)\times  \\
&&\frac{1}{\sqrt{1+{\Sigma}/{\mu} -({j_z/j_c)^2}}}
+\frac{\gamma \tilde \Delta}{4\pi(\gamma \tilde T)^3}
\frac{1}{ [1-(j_z/j_c)^2]^{3/2}} \times \nonumber \\
&&\left[\frac{1}{v_c\sqrt{1-{(j_z/j_c)^2}}}-
\left(\frac{1-v_c}{v_c}\right)\frac{1}{\sqrt{1+{\Sigma}/{\mu}-{(j_z/j_c)^2}}}
\right]^{-2}\Bigg\},  \nonumber
\end{eqnarray} 
where the parameters $v_c$ and $\Sigma$ are obtained by minimizing the
trial free energy expression given in the appendix.
This minimization is carried out numerically, and the resulting points for
$T=30^\circ K<T_{cp}\approx 35.4^\circ K$ are plotted in
Fig.~1 (c) (and can be compared to those with zero disorder
at $T=30^\circ K$ in Fig.~1 (a)).

The calculated FL fluctuations are now used to construct an
approximate `phase boundary' for the destruction of short-range
translational order. 
According to the Lindemann criterion, `melting' occurs when
$\sqrt{\left\langle\vec r\,^2\right\rangle}\approx c_L a$, 
where $c_L$ is a constant of the order of unity.
From the replica symmetric solution of Eqn.~(\ref{ICFRS3}), we obtain
a melting field
\begin{equation}\label{MTT}
\tilde B_m=c_L^4\left[\gamma\tilde T_m+\left(\frac{\gamma \tilde \Delta}{4\pi}\right)
{\left(\gamma \tilde T_m\right)^{-2}}\right]^{-2}\left(1-\frac{j_z^2}{j_c^2}\right) .
\end{equation}
In the limit of zero disorder, the above equation reproduces
the result $\tilde B_m=c_L^4(1-j_z^2/j_c^2)/(\gamma \tilde T_m)^2$
of Ref. \cite{MKMK}, and agrees with the expected behavior 
of $B\sim 1/T^2$ for zero current.
Using $c_L=0.16$, and the same parameters as in Fig.~1,
the phase diagram is plotted in Fig.~2.
For the portion of the melting line for $T<T_{cp}$, we can no longer
use the above analytical form, but must instead rely on
the numerical results obtained using Eqn.~(\ref{flu1srsb}). 
Interestingly, the melting field shows a maximum around the 
freezing temperature, even for zero current. 
Indeed, recent experimental results\cite {Khay} show 
a reduction of the melting field for low temperatures.
        
\begin{figure}
\epsfxsize=9truecm
\centerline{\epsfbox{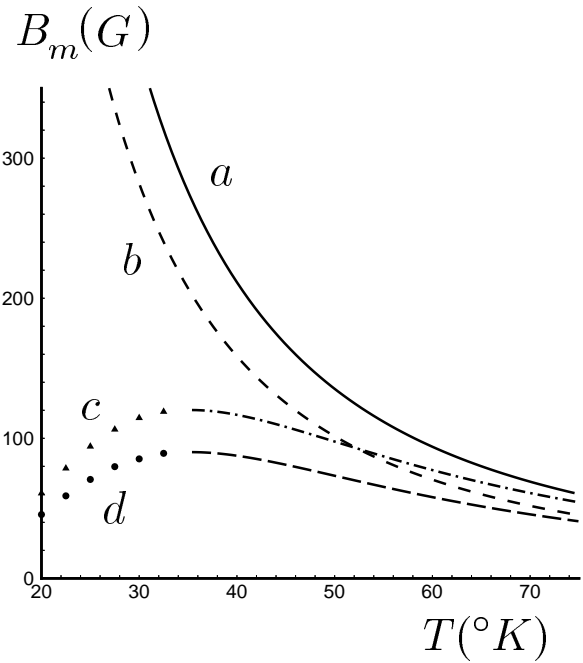}} 
\medskip
\narrowtext{\caption{`Phase diagram'
of FLs for (a) $\tilde \Delta/2\pi=0$ and $j_z/j_c=0$, (b)
$\tilde \Delta/2\pi=0$ and $j_z/j_c=0.5$, (c) 
$\tilde \Delta/2\pi=0.1$ and $j_z/j_c=0$, (d) 
$\tilde \Delta/2\pi=0.1$ and $j_z/j_c=0.5$.\label{fig2} }}
\end{figure}

\section{The Transfer matrix method}\label{tm}

With or without RSB, the variational method predicts that
in the absence of the cage potential ($\mu=0$), the FL is
unstable in the presence of infinitesimal $j_z$, and cannot be
pinned by the impurities alone.
We may still question the validity of this conclusion, given the
known shortcomings of the variational method. 
For example, in the absence of current, this method predicts
that fluctuations of a FL grow with its length $L$ as
$\left\langle\vec r\,^2\right\rangle\sim L$.
In fact, it is known that pinning to impurities results in 
$\left\langle\vec r\,^2\right\rangle\sim L^{2\zeta}$, with
$\zeta>1/2$\cite{Kardar,Zhang}.
We thus check the results using a numerically exact transfer
matrix algorithm applied to a discrete model of the FL.

Consider a FL pinned at one end to the point $(0,0,0)$ on a cubic lattice,
and denote the net Boltzmann weight of all paths terminating at
the point $(x,y,z)$ by $W(x,y,z)$.
We consider only paths that are directed along the $z$-direction,
and hence this weight can be calculated recursively as
\begin{eqnarray}\label{recrel}
&W&(z+1,x,y)=\exp\left[{-{V(z,x,y)\over T}-{\mu(x^2+y^2)\over 2T}}\right]
\times   \nonumber \\
&\Bigg\{ &W(z,x,y) +\Gamma \times \\  \nonumber
&& \Bigg[e^{y{j_z}/{j'_c}} W(z,x+1,y)
+e^{-y{j_z}/{j'_c}} W(z,x-1,y) +   \\  \nonumber
&\,\quad&e^{x{j_z}{j'_c}} W(z,x,y-1)+
e^{-x{j_z}{j'_c}} W(z,x,y+1)\Bigg]\Bigg\}, \nonumber
\end{eqnarray}
where $j'_c=2 T/\phi_0$, and $\Gamma=\exp(-\varepsilon_1/2T)$.
Note that at each step the path is either straight, or moves by only
one step in the transverse direction.
Such transverse steps acquire a weight $\Gamma\ll1$ from the 
FL elasticity, as well as a direction dependent contribution 
from the current\cite{Medina}.

\begin{figure}
\epsfxsize=6.7truecm
\centerline{\epsfbox{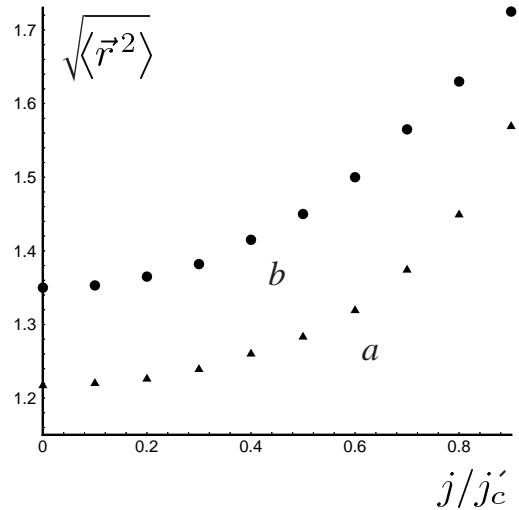}} 
\medskip
\narrowtext{\caption{The mean-square fluctuations of a FL with $\Gamma=1$,
confined to a cage of  $\mu/T=0.2$, as a function of the reduced
 current $j_z/j'_c$, for (a) zero disorder; and (b)  binary random potentials 
chosen from $(0.2,2)$.
\label{fig3} }}
\end{figure}

Starting with the initial values of $W(0,x,y)=\delta_{0x}\delta_{0y}$,
successive applications of this recursion relation provide
the weights $W(x,y,z)$ at subsequent layers, which can then be used to calculate the
(exact) degree of fluctuations 
$\left\langle \vec r\,^2\right\rangle=\left\langle x^2+y^2\right\rangle$, 
for a particular realization of the random potential $V(x,y,z)$. 
The results are then averaged over different realizations of the
disorder potential.
When the FL is confined to a cage ($\mu\neq 0$), 
$\left\langle \vec r\,^2\right\rangle$ quickly saturates to a constant value.
Fig.~\ref{fig3} shows the dependence of the saturation point
on the external current.
For this figure, the random potentials $V/T$ were taken from a binary set 
$(0.2,2)$, and the results averaged over $500$ realization of  randomness.
Since the microscopic models are different, it is hard to make
a quantitative comparison with the replica calculations.
The numerical results are certainly in qualitative agreement with
the replica estimates depicted in Fig.~\ref{fig1}.
It is numerically difficult to probe currents close to the critical value,
as the large transverse excursions require simulating large lattices.

\begin{figure}
\epsfxsize=8truecm
\centerline{\epsfbox{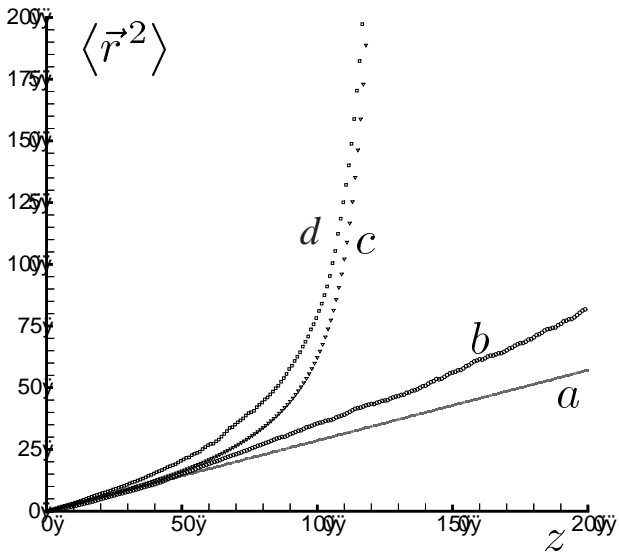}} 
\medskip
\narrowtext{\caption{Mean-square fluctuations of a
FL in a $200\times 50\times 50$ lattice, with (a)
$V=0, j_z/j_c=0$ (b) $V\neq 0, j_z/j_c=0$ (c) $ V=0, j_z/j_c=0.08$,
and (d) $V\neq 0, j_z/j_c=0.08$.
\label{fig4} }}
\end{figure}

The most interesting point  is the behavior of the FL in the absence 
of the cage potential.
As can be seen from the various limiting curves for $\mu=0$,
plotted in Fig.~\ref{fig4}, fluctuations no longer saturate, but 
continue to grow with $z$.
When both the current and disorder are zero, the FL behaves as
a directed random walk.
The straight line in Fig.~\ref{fig4}(a) corresponds to 
$\left\langle\vec r\,^2\right\rangle\propto t$, as expected in this case.
The addition of point disorder changes the scaling of transverse
fluctuations\cite{Kardar}. 
As indicated in Fig.~\ref{fig4}(b), the average of
$\left\langle\vec r\,^2\right\rangle$ grows faster than $t$.
The results for the binary set $(\{0.2,2\})$ of random potentials $V/T$,
averaged over  $500$ realization of randomness, are consistent with 
$\overline{\left\langle\vec r\,^2\right\rangle}\propto t^{2\zeta}$,
with $\zeta\approx 0.6$. This is in agreement with previous results
for directed polymers in random media \cite{Kardar}.
As indicated in Fig.~\ref{fig4}(c) the addition of current to
the pure system ($V=0$) leads to instability, and the transverse
fluctuations become unbounded.
When both binary disorder and current act on the FL, it is again
found to be unstable, as depicted in Fig.~\ref{fig4}(d).
Indeed, we find that even an infinitesimal current can 
destabilize the FL, and that binary impurities only further
increase transverse fluctuations.

\begin{figure}
\epsfxsize=8truecm
\centerline{\epsfbox{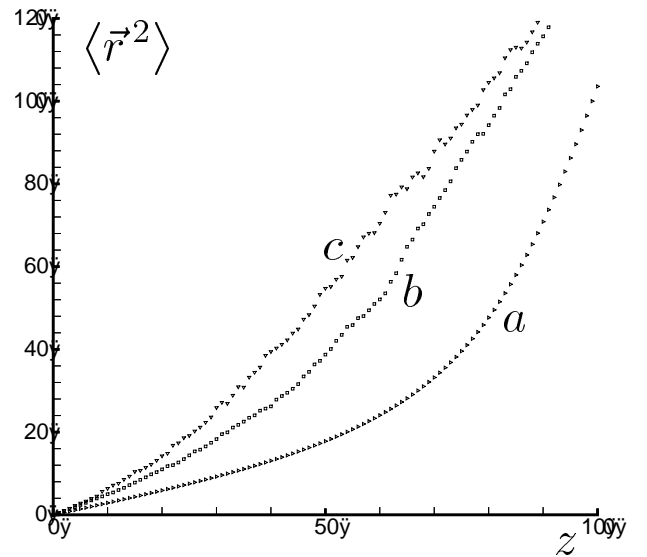}} 
\medskip
\narrowtext{\caption{Transverse fluctuations for
$j_z/j_c=0.1$, with $500$ realizations of the random potential
$V/T$ taken from a Gaussian distribution of
variance (a) $0$, (b) $2$, and (c) $4$.
\label{gauss} }}
\end{figure}

Since the use of binary randomness in directed polymer simulations
may lead to anomalous results related to percolation\cite{Chate},
we also repeated the simulations using a  Gaussian distribution 
for the disorder potential $V/T$.  Such impurities are more effective
in pinning the FL, and appear to stabilize it at small scales.
However, by varying the width of the lattice or $z$, we
see that the FL goes to the edges of the box, indicating its instability.
Fig.~\ref{gauss} depicts the dependence of transverse fluctuations
on the strength of Gaussian disorder.
The  fluctuations at short scales are governed by disorder, while
at longer scales the current dominates, pushing the FL outward.

\section{Onset of the Instability}\label{onset}

The results of replica and transfer matrix computations can be summarized as follows:
A parallel current $j_z$ causes instabilities in a flux lattice, or in a single (unconfined) FL.
Point impurities appear to only increase the extent of fluctuations, and do not
provide stability.
On the other hand, if we start with a vortex system pinned by strong point
impurities at low temperatures, it is difficult to imagine how an infinitesimal
$j_z$ can lead to drastic changes in configuration.
In this section we discuss the length scales, barriers, and time scales that
control the onset of the instability.

Let us first consider a single FL in the absence of point impurities or a
confining potential. 
A normal mode of wavenumber $q_z$ has energy proportional
to $G_0^{-1,L/R}(q_z)=\varepsilon_1 q_z^2\pm\phi_0 j_z q_z$, indicating a band of
unstable modes for $q_z\leq \phi_0 j_z/\varepsilon_1$.
The instability occurs only on segments of length longer
than $\ell^*\sim \varepsilon_1/(\phi_0 j_z)$.
(The divergence of $\ell^*$ as $j_z\to0$ is cut-off by a finite confining potential.)
The simplest equation of motion for the FL relates its local velocity to the
force obtained from variations of the Hamiltonian by
$d\vec r/dt=-\hat\mu[\delta{\cal H}/\delta \vec r\,]$, where $\hat\mu$
is a mobility\cite{MKMK}.
The characteristic time scale for the normal modes is now given by
$\tau^{-1,L/R}(q_z)=\hat\mu G_0^{-1,L/R}(q_z) $.
The time required for the growth of unstable modes diverges as
$\tau^*\sim \hat\mu\varepsilon_1^2/(\phi_0 j_z)^2 $ for $j_z\to 0$.
The FL thus appears stable at short times, or  at high frequencies $\omega$.

The corresponding analysis for a flux lattice in the absence of impurities
is very similar. The (inverse) time scales for growth or decay of normal modes
are related to the corresponding propagators $G_0^{-1,L/R}$ in
Eqn.~(\ref{DecomG}) by a mobility factor.
The unstable modes occur over an approximately spherical 
region centered on $q_z\leq \phi_0 j_z/2c$,
and the time scale for the onset of instability scales as
$\tau^*\sim \hat\mu c^2/(\phi_0 j_z)^2$.

In the presence of point impurities, we claim that the instability
still occurs for length scales larger than $\ell^*\sim (\phi_0 j_z)^{-1}$.
The analytical support for this claim comes from the replica analysis:
The best variational propagators (as for example in Eqn.~(\ref{FinG2}))
are proportional to the pure one with the same range of instabilities.
There is also numerical support from the transfer matrix calculations:
The comparison of Figures 4(c) and 4(d) indicates that the
onset of instability is not substantially modified in the presence
of the disorder potential.
The location of the instability (as determined from such curves for
$\left\langle \vec r\,^2\right\rangle$) is plotted in  Fig.~6,
and indeed scales as $z^*\sim 1/(j_z/j'_c)$.

\begin{figure}
\epsfxsize=8truecm
\centerline{\epsfbox{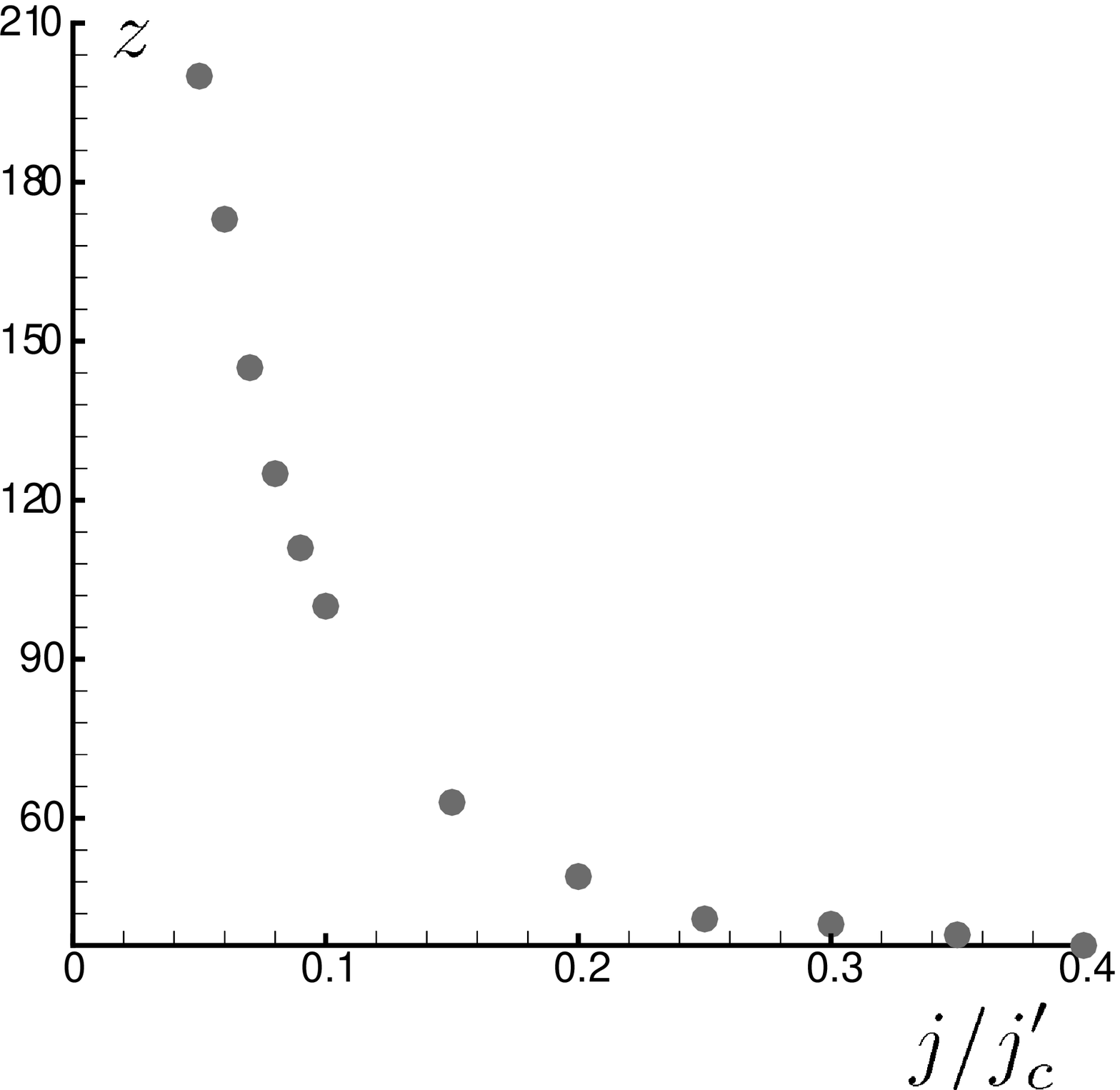}} 
\end{figure}
{\small {Fig.~6: The behavior of the position for the onset of instability
as a function of  the current.}}

In response to the current, the unstable FLs must change configuration
at a length scale $\ell^*$.
This is not easily achieved in the presence of collective pinning by
point impurities, and the flux system requires the assistance of
thermal fluctuations to overcome the pinning barriers\cite{Blatter,FFH}.
For a single FL the strength of the collective barriers grows with the
length scale as $E_b\propto \ell^\theta$ with 
$\theta\approx0.3$\cite{Drossel}.
As $j_z\to0$ the barriers at the length scale for the onset of instability
thus diverge as $E_b\propto j_z^{-\mu}$ with $\mu=\theta$.
The probability that the flux system has sufficient thermal energy
to overcome such barriers is proportional to the Boltzmann 
weight $\exp(-{E_b/k_BT})$.
The waiting time for such rare events grows as
$\tau^*\sim \tau_0\exp[(j_T/j_z)^\mu]$, where $\tau_0$ is
a microscopic time scale, and $j_T$ is a characteristic current
scaling as $T^{-1/\mu}$.
   
The above arguments  essentially follow those of Ref.\cite{FFH}
which considers the response of a  collectively pinned flux system
to a current perpendicular to the magnetic field.
They are equally valid for a collection of flux lines in the vortex glass
or Bragg glass phase.
The  exponent $\theta$ depends on the glass phase considered,
its exact value is not as well known as for a single FL.
Thus while the equilibrium state of a flux system is destabilized by an 
infinitesimal $j_z$, the time for the onset of such instability diverges.
The instability will not be observed if the system is probed at times
shorter than $\tau^*$, or perturbed by an AC current $j_z(\omega)$
with frequency $\omega>\omega^*\sim 1/\tau^*$.
(The latter is in qualitative agreement with the phase diagram
presented in Ref.\cite{MKMK}.)

\acknowledgements
We have benefited from discussions with J. Davoudi, Y. Goldschmidt,
R. Golestanian, and M. Mezard. 
M. Kohandel acknowledges support from IASBS, Zanjan, Iran,
and IPM, Tehran, Iran.
M. Kardar is supported by the National
Science Foundation (Grant No.  DMR-98-05833).

\appendix
\section{Replica Symmetry Broken solutions for a single FL}\label{RSB}
In order to find the replica symmetry breaking (RSB) solutions, 
we parameterize $\sigma_{ab}$ by a function $\sigma(v)$ 
defined for $0\leq v\leq 1$. The propagators 
$G_{ab}(q_z)$ and $G_{aa}(q_z)$ are also replaced by $G(q_z,v)$ and
$\tilde G(q_z)$, respectively. 
The  stationarity equations are now written as 
\begin{equation}\label{sigmav1s}
\sigma(v)=\frac{2}{T}\tilde f_p'\left[\frac{1}{2}\left(B^L(v)+B^R(v)\right)\right],
\end{equation}
where
\begin{equation}\label{BLTv1s}
B^{L/R}(v)=2T\int\frac{dq_z}{2\pi}\left[\tilde G^{L/R}(q_z)-
G^{L/R}(q_z,v)\right].
\end{equation}
Inversion of the hierarchical matrices gives \cite{MezPar},
\begin{eqnarray}\label{invG1s}
&&\tilde G^{L/R}(q_z)-G^{L/R}(q_z,v)=\\
&&\frac{1}{v\left(G_0^{-1,L/R}(q_z)+[\sigma](v)\right)}
-\int_v^1 \frac{d\omega}{\omega^2}
\frac{1}{G_0^{-1,L/R}(q_z)+[\sigma](\omega)},  \nonumber
\end{eqnarray}
where $[\sigma](v)=v\sigma(v)-\int_0^vd\omega\sigma(\omega)$  as before.
Integration over $q_z$ now leads to 
\begin{equation}\label{siggv}
\sigma(v)=\frac{2}{T}\tilde f_p'\left[\frac{1}{v}g\left([\sigma](v)\right)-
\int_v^1 \frac{d\omega}{\omega^2}g\left([\sigma](\omega)\right)\right],
\end{equation}
where $g\left([\sigma]\right)=
T/\sqrt{4\varepsilon_1\left(\mu+[\sigma]\right)-\phi_0^2 j_z^2}$. 

For a solution with 1-step RSB,  we assume\cite{MezPar,Gold}, 
\begin{equation}\label{sig1st}
\sigma(v)=\left\{
\begin{array}{ll}
\sigma_0 & {\rm for}~v< v_c ,\\
\sigma_1 & {\rm for}~v\geq v_c .
\end{array}
\right.
\end{equation}
This yields $[\sigma](v)=0$ for $v\leq v_c$,
and $[\sigma](v)=\Sigma=v_c(\sigma_1-\sigma_0)$ for $v\geq v_c$. 
We can substitute the above form in Eqn.~(\ref{invG1s})
and use the result to evaluate $B^{L/R}$ in Eqn.~(\ref{BLTv1s}).
Depending on whether $v$ is bigger or smaller than $v_c$, we
obtain two different values for $B^{L/R}$, which when
placed in Eqn.~(\ref{sigmav1s}), lead to the self-consistency
conditions
\begin{eqnarray}\label{sigma01s}
\sigma_0&=&\frac{2}{T} \tilde f_p'\left[\frac{2T}{v_c}\frac{1}{s_2}
-\frac{2T(1-v_c)}{v_c}\frac{1}{s_1}\right] \\  
\sigma_1&=&\frac{2}{T} \tilde f_p'\left[\frac{2T}{s_1}\right],  \nonumber
\end{eqnarray}
where $s_1=\sqrt{4\varepsilon_1(\mu+\Sigma)-\phi_0^2j_z^2}$  
and $s_2=\sqrt{4\varepsilon_1\mu-\phi_0^2j_z^2}$
(for $j_z<\sqrt{4\varepsilon_1(\mu+\Sigma)/\phi_0^2}$).
One can evaluate the free energy as a function of the parameters
$\Sigma$ and $v_c$ of the 1-step RSB solution, as
\begin{eqnarray}\label{F1sRSB}
&&\frac{F}{2L}={\rm const.}+\frac{\tau}{4}\frac{1-v_c}{v_c}\Bigg\{
\Sigma\left[1+\frac{\Sigma}{\mu}-\frac{j_z^2}{j_c^2}\right]^{-1/2}-\\
&&2\mu\left[1+\frac{\Sigma}{\mu}-\frac{j_z^2}{j_c^2}\right]^{1/2}\Bigg\}
-\frac{1-v_c}{2T}\tilde f_p\left(\frac{\tau}{\sqrt{1
+{\Sigma}/{\mu}-{j_z^2}/{j_c^2}}}\right) \nonumber \\
&&-\frac{v_c}{2T}\tilde f_p\left(\frac{\tau}{v_c\sqrt{1
-{j_z^2}/{j_c^2}}}-\frac{1-v_c}{v_c}\frac{\tau}{\sqrt{1
+{\Sigma}/{\mu}-{j_z^2}/{j_c^2}}}\right), \nonumber
\end{eqnarray}
where $\tau=T/\sqrt{\mu\varepsilon_1}$. 
Variation with respect to $\Sigma$ reproduces the condition
$\Sigma=v_c(\sigma_1-\sigma_0)$, $\sigma_0$ and $\sigma_1$
defined in Eqn.~(\ref{sigma01s}).
Minimizing the free energy with respect to the $v_c$, 
after some algebraic manipulations, leads to
\begin{equation} \label{vc1s}
v_c=\frac{\tilde T}{\tilde T_{cp}},
\end{equation} 
with $\tilde T_{cp}=(\tilde \Delta/2\pi\gamma^2)^{1/3}$. 
Since $v_c\leq 1$, this solution exists only for temperatures
lower than a freezing temperature, which 
{\em is independent of the current and $\mu$}. 
With this value of $v_c$ inserted in Eqs.~(\ref{sigma01s}),  
we can numerically find $\Sigma$ for any temperature
and current, and evaluate the FL fluctuations in Eqn.~(\ref{flu1srsb}).

The natural question is whether there are also solutions with 2-step, 
or even full RSB.
Taking a derivative of Eqn~(\ref{siggv}) with respect to $\sigma$, 
and using the functional form for $\tilde f_p$, 
leads after further algebraic manipulations,
to a linear relation between $\sigma(v)$ and $[\sigma](v)$.
This is inconsistent with the relation 
$[\sigma](v)=v\sigma(v)-\int_0^vd\omega\sigma(\omega)$,
unless $\sigma$ is a constant
(since $\sigma'(v)=0$ is always a solution).
One the other hand, one can also show by taking two derivatives 
with respect to $v$,  that the only possible solution
is $v=v_c=\tilde T/\tilde T_{cp}$, consistent with the 1-step RSB result.
We thus conclude that, at least for this particular form for the disorder
correlator (and hence $\tilde f_p$), there is no solution 
with continuous RSB.

\end{multicols}

\begin{references}
\bibitem{FotN}
When the parallel penetration depth $\lambda_{\parallel}$ is finite,
our result are applicable only to thin samples with thickness
of a few $\lambda_{\parallel}$, or alternatively,
only close to the surface of the sample.
\end{references}
\end{document}